\documentclass[12pt]{article}
\newcommand{\fr}{\frac}
\newcommand{\lb}{\label}
\newcommand{\ti}{\tilde}
\newcommand{\be}{\begin{equation}}
\newcommand{\ee}{\end{equation}}
\newcommand{\ba}{\begin{array}}
\newcommand{\ea}{\end{array}}
\newcommand{\beqa}{\begin{eqnarray}}
\newcommand{\beqaa}{\begin{eqnarray*}}
\newcommand{\al}{\alpha}
\newcommand{\ald}{{\dot{\alpha}}}
\newcommand{\bet}{\beta}
\newcommand{\betd}{{\dot{\beta}}}

\newcommand{\si}{\sigma}

\newcommand{\sib}{{\bar{\sigma}}}
\newcommand{\te}{\theta}
\newcommand{\teb}{{\bar{\theta}}}
\newcommand{\chib}{{\bar{\chi}}}
\newcommand{\del}{\partial}
\newcommand{\eeqa}{\end{eqnarray}}
\newcommand{\eeqaa}{\end{eqnarray*}}
\newcommand{\ep}{\epsilon}

\newcommand{\Dc}{{\cal D}}

\newcommand{\Oc}{{\cal O}}

\newcommand{\omo}{\omega}
\newcommand{\etat}{\tilde{\eta}}
\newcommand{\omob}{{\bar{\omega}}}

\newcommand{\CD}{\Delta}

\newcommand{\rD}{{\rm D}}

\begin{document}
\title{}
\author{}
\date{}

\noindent
{\Large \bf {$\mathbf{N=1}$ 
 supersymmetric Yang--Mills theory
in $\mathbf{d=4}$
and its Batalin--Vilkovisky  
quantization  by spinor \mbox{superfields}}}

\vspace{2cm}
\noindent
 \"{O}mer F. DAYI\footnote{E-mail addresses: dayi@itu.edu.tr, dayi@gursey.gov.tr.} 

\vspace{10pt}

\noindent
{\footnotesize {\it Physics Department, Faculty of Science and
Letters, Istanbul Technical University,
TR--34469  Maslak--Istanbul, Turkey. } }

\vspace{10pt}

\noindent
{\footnotesize \it {Feza G\"{u}rsey Institute,
P.O.Box 6, TR--34684
\c{C}engelk\"{o}y--Istanbul, Turkey. }}

\vspace{3cm}

Four dimensional 
$N=1$ supersymmetric Yang--Mills theory action 
is written  in terms of the
spinor superfields in transverse 
gauge. This action is seemingly first order in 
space--time derivatives. Thus, it suggests that the
generalized fields approach of obtaining 
Batalin--Vilkovisky quantization can be applicable. 
In fact, 
generalized fields which collect spinor superfields
possessing different ghost numbers 
are introduced to obtain the minimal 
solution of its Batalin--Vilkovisky master
equation in a compact form.

\newpage

\section{Introduction}

One of the important elements in the Berkovits
formulation\cite{ber}
 of covariant quantization 
of ten dimensional  relativistic  superparticle and superstring
theories  is the
spinor superfield which was  introduced in \cite{hs}.
Being a superconnection component, this spinor  superfield
 possesses gauge freedom.
In fact, 
its component fields are derived
in the so called transverse gauge, where 
the gauge freedom  depending 
on anticommuting coordinates is eliminated. 
Transverse gauge was first  introduced  
in \cite{hhls} 
to study some aspects of $N=3$
supersymmetric Yang--Mills theory
in four dimensional superspace. We  
focus on the spinor superconnection components 
given in \cite{hhls}
to deal with $N=1$ 
supersymmetric  Yang--Mills  theory  in $d=4.$
We show that  spinor superconnection in transverse gauge 
can be  used in an action which
yields
$N=1$ super Yang--Mills  action in component fields once the integrations 
over  anticommuting coordinates of the superspace
are performed.
This action is 
apparently linear in space--time derivatives.

A formalism for  quantization of  gauge theories 
respecting the original symmetries is due to 
Batalin and Vilkovisky (BV)\cite{bv}.
It is a systematic procedure of obtaining actions of gauge theories
which can be used in the related path integrals. 
This method is formulated in terms of ghost fields and antifields.
Utilizing these fields 
BV--master equation  is introduced 
based on the
 BRST invariance of quantized gauge theory actions.
Solution of BV--master equation is the action which can be
employed  in the related path integrals. 
To have a better understanding of different 
features of BV--method
the initial  gauge fields, ghosts and their
antifields are gathered as components of 
generalized fields\cite{o1} or 
superfields\cite{tsup}\footnote{For some 
applications and similar formalisms
see the references in \cite{o2}.}. 
We denote the latter as $\tau$--fields for 
distinguishing them from
the other  superfields. It was shown that
these formalisms are related\cite{o2}:
$\tau$--fields can be seen as a systematic way of writing 
generalized fields and their products. 
In terms of these formalisms mostly gauge theories
which are not supersymmetric are considered with
an exception presented in \cite{o2}.
Writing four dimensional $N=1$ supersymmetric Yang--Mills action
in terms of spinor superfields
in transverse gauge
suggests that a generalized spinor
superfield can be introduced to obtain 
the minimal solution of its
BV--master equation.
This compact form of the 
minimal solution of BV--master equation
can be useful
to understand geometrical aspects of the theory.
Moreover, it  constitutes an example of applicability 
of the generalized fields and $\tau$--fields methods 
to theories which are
second order in space--time derivatives, although the
original form of the action 
is seemingly first order in the derivatives.

In the next section we start with  recalling 
definitions of superconnection 
components of  
four dimensional $N=1$
supersymmetric Yang--Mills theory 
in transverse gauge. Afterward,
we employ them in
an  action whose kinetic term
is bilinear in the spinor superfields 
and  first order  in  derivatives.
Its interaction part is given with superconnection
components. 
We  show that
after integrating over anticommuting variables
$N=1$ supersymmetric Yang--Mills action in component
fields follows. 
In Section 3 we introduce ghost fields and
antifields to write 
 the appropriate generalized superfields or 
equivalently $\tau$--fields.
By generalizing the action of Section 2 we show that
BV--quantization can be achieved in a compact form
utilizing these generalized fields.
Lastly we discuss the results obtained.

\section{Formulation  of 
$\mathbf{N=1}$ 
supersymmetric \mbox{Yang--Mills} action by spinor superfields}

We deal with the four dimensional superspace 
whose coordinates are 
$( x^{\al \ald},\ \te^\al ,\teb^\ald ).$
$(\al,\ald )$ are the two component spinor 
indices\footnote{Conventions of \cite{wb} are used and for 
explicit calculations in superspace see \cite{mw}. }.
Let us 
briefly mention the procedure presented in
\cite{hhls} to find  the 
superconnection components
$\omo_\al ,\omob_\ald ,\ti{A}_{\al \ald},$
starting from the gauge  and spinor fields
$A_{\al \ald}, \chi_\al , \chib_\ald ,$
in the transverse gauge:
\be
\lb{tgc}
\te^\al\omo_\al +\teb^\ald \omob_\ald =0.
\ee
Superconnection components
are related by some constraint equations
and Bianchi identities. Hence, they are
defined to satisfy some differential equations.
We deal with $N=1$ supersymmetric Yang--Mills
theory, so that,  
one should solve the following equations given in terms 
of the superfields
$\ti{A}_{\al\ald} ,\ti{\chib}_\ald ,\ti{\chi}_\al ,$
\beqa
\Dc \ti{A}_{\al\ald} & = & -i\te_\al \ti{\chib}_\ald 
-i\teb_\ald \ti{\chi}_\al , \lb{de1}\\
\Dc \ti{\chib}_\ald &=& 2 \teb^\betd \ti{F}_{\ald\betd} ,\\
\Dc \ti{\chi}_\al &=& 2 \te^\bet \ti{F}_{\al\bet} , \lb{de3}
\eeqa
where we introduced   the operator
$$
\Dc=\te^\al\fr{\del}{\del\te^\al} +
\teb_\ald\fr{\del}{\del\teb_\ald}.
$$ 
$\ti{A}_{\al\ald} ,\ti{\chib}_\ald ,\ti{\chi}_\al $
are the superfields whose first components are 
$A_{\al\ald} ,\sqrt{3}\chib_\ald ,\sqrt{3}\chi_\al .$
All of them take values in the gauge algebra.
Here we adopt  a normalization which is 
suitable to our formalism.
Selfdual and anti-selfdual  field strengths
are given, respectively,  as
\beqa
\ti{F}^{\al\bet} & =& 
\ep_{\ald \betd}
\left[ 
(\del^{ \ald \al }+i\ti{A}^{ \ald \al}) \ti{A}^{\betd \bet} 
-(\del^{ \betd \bet }+i\ti{A}^{ \betd \bet}) \ti{A}^{\ald \al}\right] ,\\
\ti{F}^{\ald\betd} & = & 
\ep_{\al \bet}\left[ 
(\del^{\ald \al } +i\ti{A}^{\ald \al})\ti{A}^{\betd \bet} 
-(\del^{\betd \bet } +i\ti{A}^{\betd \bet})\ti{A}^{\ald \al}) \right].
\eeqa
Once $\ti{A}_{\al\ald} $ is obtained,
the spinor superconnections $\omo_\al ,\omob_\ald $ 
can be derived by  the equations
\beqa
(1+\Dc )\omo_\al & = & 2\teb^\ald \ti{A}_{\al\ald} \lb{om} , \\
(1+\Dc )\omob_\ald & = & 2\te^\al \ti{A}_{\al\ald} \lb{omb} .
\eeqa
By solving (\ref{de1})--(\ref{de3}) for
$\ti{\chib}_\ald ,\ti{\chi}_\al $
and $\ti{A}_{\al\ald}, $
up to the second order in $\te_\al ,\teb_\ald , $ 
one finds
\beqa
\ti{A}_{\al\ald} & = & A_{\al\ald}
-i\sqrt{3}\te_\al \chib_\ald  -i\sqrt{3}\teb_\ald \chi_\al  
-i\te_\al\teb^\betd F_{\ald\betd}
+i\te^\bet\teb_\ald F_{\al\bet} +\cdots , \lb{at} \\
{\ti{\chi}}_\al & = &\sqrt{3} \chi_\al +2 \te^\bet F_{\al\bet}
-i\frac{3 \sqrt{3}}{2}\te^2 \rD_{\al\ald}\chib^\ald
-i \sqrt{3}\te^\bet \teb^\ald \rD_{\al \ald} \chi_\bet \nonumber \\
& &  
-i \sqrt{3}\te^\bet \teb^\betd \rD_{\bet \betd} \chi_\al  + \cdots , \\
{\ti{\chib}}_\ald & = & \sqrt{3}\chi_\ald +2 \teb^\betd F_{\ald\betd}
+i \frac{3\sqrt{3}}{2}\teb^2 \rD_{\al\ald}\chi^\al
+i \sqrt{3}\te^\bet \teb^\betd \rD_{\bet \betd } \chib_\ald  
\nonumber \\
& &  
+i \sqrt{3}\te^\al \teb^\betd \rD_{\al \ald } \chib_\betd  
+ \cdots , 
\eeqa
in terms of
the covariant derivative 
$\rD_{\al\ald}=
\del_{\al \ald}+i[A_{\al \ald},].$ 
Now, by solving (\ref{om}) and (\ref{omb})
we can write the spinor superconnection components  as
\beqa
\omo_\al & = & \teb^\ald A_{\al\ald} +\fr{2i}{\sqrt{3}}\teb^2\chi_\al 
+\fr{2i}{\sqrt{3}}\ep_{\al \bet}\te^\bet \teb^\betd \chib_\betd 
+\fr{i}{2}\teb^2\te^\bet F_{\bet\al}
+\fr{\sqrt{3}}{5}
\te^2\teb^2 \rD_{\al\ald}\chib^\ald , \lb{omo}\\
\omob_\ald &  = & \te^\al A_{\al\ald} -\fr{2i}{\sqrt{3}}\te^2\chib_\ald 
-\fr{2i}{\sqrt{3}}\ep_{\ald \betd}\te^\bet \teb^\betd \chi_\bet 
 -\fr{i}{2}\te^2\teb^\betd F_{\ald\betd}
-\fr{\sqrt{3}}{5}
\te^2\teb^2\rD_{\al\ald}\chi^\al . \lb{omob}
\eeqa

We would like to utilize the superconnection components
(\ref{at}),(\ref{omo}) and (\ref{omob})
in transverse gauge (\ref{tgc})
to obtain $N=1$ supersymmetric Yang--Mills theory action.
To achieve this 
let us introduce
\be
\lb{delta}
\CD_{\al \ald}\equiv \del_{\al \ald} +\fr{i}{4}[\ti{A}_{\al \ald} ,]
\ee
and define 
the action
\be
\lb{soA}
S_0 = 
-\fr{i}{2}< \omo^\al \CD_{\al \ald}\omob^\ald > 
-\fr{i}{2}< \omob_\ald \CD^{\ald \al}\omo_\al >,
\ee
where we  suppress trace over the gauge algebra and
adopt the notation
\[
<\Oc >\equiv \int d^4x d^2\te d^2\teb \  \Oc.
\]
$ [,]$ is the generalized
commutator: Anticommutator when both of the entries are
fermionic, commutator otherwise.
Integrals  over the anticommuting variables vanish except
\beqa
\int d^2 \te \ \te^\al \te^\beta & = &\fr{1}{2} \ep^{\beta \al}, \nonumber \\
\int d^2 \teb \ \te^\ald \te^\betd & = & \fr{1}{2}\ep^{\ald \betd }. \nonumber
\eeqa
Performing the integrals over the anticommuting coordinates 
$\te_\al ,\teb_\ald $
in (\ref{soA}) one reaches to 
\be
\lb{soo}
S_0 = \fr{-1}{2}\int d^4x\ \left[\fr{1}{4}F^{\al\bet}F_{\al\bet}
+\fr{1}{4}F_{\ald\betd}F^{\ald\betd} 
+i\chi^\al \rD_{\al\ald}\chib^\ald
+i\chib_\ald \rD^{\ald\al}\chi_\al \right] .
\ee
This is the $N=1$ supersymmetric Yang--Mills
theory  action without auxiliary fields. 

\section{The BV--quantized  action  by generalized spinor superfields}

Form of the action (\ref{soA}) which is seemingly first order in derivatives 
advocate to accomplish its BV quantization by
 the generalized fields approach\cite{o1} or equivalently
the $\tau$--fields  method\cite{tsup}.
One deals with  kinetic
and interaction parts of the actions separately
 in the original formulations of these methods.
However  here, a unified
treatment is preferred.

Gauge transformations of $N=1$ supersymmetric Yang--Mills theory 
are irreducible. Therefore, to perform its
quantization in a covariant manner
we introduce the anticommuting,
gauge algebra valued
 ghost field $\eta .$ 
To acquire $\tau$--fields
let us also introduce the Grassmann variables
$\tau_\mu ,$ where $\mu$ is  4 dimensional vector index.
$\tau_\mu$
 are  defined to possess ghost
number 1 as the ghost field $\eta .$  
Then,  using   
$\CD_\mu =\del_\mu +(i/4)[\ti{A}_\mu ,]$ we 
introduce the operator
\be
\lb{bvso}
\CD_\tau = \tau_\mu \CD^\mu +i[\ti{\eta} ,] 
\ee
possessing ghost number 1. The ghost number 1 superfield 
$\etat ,$ whose component fields will be specified later, 
is associated to the ghost field $\eta .$
 As it is announced the operator (\ref{bvso}) is acquainted 
with interactions.
One  introduces  generalized spinor superfields which
are the $\tau$--fields 
carrying  spinor index, 
$\Psi_{\tau\al}$ and $\Phi_{\tau \al} ,$ 
to write the action  
\be
\lb{sma}
S_1=\fr{i}{2} < \int d^4\tau 
\Psi_\tau^\al 
\CD_\tau 
\Phi_{\tau \al} >.
\ee
Trace over the gauge algebra is suppressed.
A physical
action is defined to possess  ghost number zero
and even Grassmann parity.
 Thus,
due the ghost number attributed to $\tau_\mu $ we
choose 
$\Psi_{\tau \al}$ and $\Phi_{\tau\al}$ to have
ghost numbers 0 and 3, respectively.
Moreover, we let $\Psi_{\tau \al}$ to be Grassmann odd
and $\Phi_{\tau\al}$ to be Grassmann even.
A consistent choice for the components of $\tau$--fields
(generalized fields) is
\beqa
\Psi_{\tau \al} & = & \Psi_{0\al}+\tau_\mu \Psi_{1\al}^\mu 
+\tau_\mu \tau_\nu \Psi_{2\al}^{\mu\nu} , \lb{tp1}\\
\Phi_{\tau \al} & = & 
\tau_\mu \tau_\nu \Phi_{2\al}^{\mu\nu} +
\tau_\mu \tau_\nu \tau_\rho 
\Phi_{3\al}^{\mu \nu \rho}
+ \tau_\mu \tau_\nu \tau_\rho \tau_\si 
\Phi_{4\al}^{\mu \nu \rho \si}. \lb{tp2}
\eeqa
Ghost number $N_g$ and  Grassmann parity $\ep$ 
of the components of $\Psi_\tau$ and $\Phi_\tau$ 
are 
\[
\begin{array}{rrrrrrrr}
& \Psi_0 &  \Psi_1 &  \Psi_2  & &
  \Phi_2 & 
 \Phi_3 &  \Phi_4  \\ 
N_g & 0 & -1 & -2  & &
 1 & 0 & -1 \\
\ep & 1 & 0 & 1  & &
 0 & 1 & 0
\end{array}
\]
Negative ghost number fields are antifields of the BV formalism.

Without altering the number of linearly independent elements
and taking into consideration ghost numbers we  define
\beqa
\Phi_{4\al}^{\mu \nu \rho \si} & = & 
\fr{-1}{4!}
\ep^{\mu \nu \rho \si} \Psi^\star_{0\al}, \\
\Phi_{3\al}^{\mu \nu \rho} & =  &
\fr{1}{3!}
\ep^{\mu \nu \rho \si} \Phi_{1\al\si} ,\\
\Phi_{2\al}^{\mu\nu} &= &  \ep^{\mu \nu \rho \si}
\phi_{2\al \rho\si} ,\\
\Psi_{1\al}^\mu & = &\fr{1}{2}\Phi_{1\al}^{\star \mu},
\eeqa
where star indicates the antifield as usual. 

$\tau$ integrations vanish except the following one
\be
\int d^4\tau \ \tau^\mu \tau^\nu  \tau^\rho  \tau^\si = 
\ep^{\mu \nu \rho \si} .
\ee

By plugging the generalized fields
(\ref{tp1}),(\ref{tp2}) into the action (\ref{sma})
and integrating over $\tau_\mu$ one finds
\be
S_1  = \fr{1}{2} <-i \Psi_{0}^{\al}  \CD_\mu \Phi_{1\al}^{\mu} 
+\Psi_{0}^{\al} [ \etat , \Psi^{\star}_{0 \al}]
+i\Phi_{1}^{\star \al \mu}  \CD^\nu \phi_{2 \al\mu \nu} 
-\fr{1}{2}\Phi_{1}^{\star \al \mu}[\etat , \Phi_{1 \al\mu}] 
 -4\Psi_{2}^{\al\mu\nu} [\etat ,  \phi_{2 \al \mu \nu}]> .\lb{spsi1}
\ee

Similarly, we introduce the action for the $\tau$--fields 
(generalized fields) with dotted indices:
\be
\lb{smad}
S_2= \fr{i}{2} < \int d^4\tau 
\Psi_{\tau\ald} 
\CD_\tau 
\Phi_\tau^\ald >.
\ee
Component fields of
$\Psi_{\tau\ald} $ and 
$\Phi_\tau^\ald $
are chosen in accord with to the undotted 
ones (\ref{tp1}),(\ref{tp2}). Hence, 
after performing the $\tau_\mu$ integrations 
in (\ref{smad}) one gets
\be
\lb{spsi2}
S_2= \fr{1}{2}<- i\Psi_{0\ald}  \CD_\mu \Phi_{1}^{\ald \mu} 
+\Psi_{0\ald} [ \etat , \Psi_0^{\star  \ald} ]
+ i\Phi_{1\ald}^{\star \mu}  \CD^\nu \phi_{2 \mu \nu}^\ald 
-\fr{1}{2}\Phi_{1\ald }^{\star\mu} [\etat , \Phi_{1 \mu}^\ald ]
-4\Psi_{2\ald}^{\mu\nu}  [\etat , \phi_{2 \mu \nu}^\ald ] > .
\ee

To express the component fields in terms of spinor superfields
we adopt the definitions
\beqa
\Phi_{1\al}^{\mu}   =  \si_{\al \ald}^\mu \Phi_0^\ald , &
\Phi_{1\mu}^{\ald}    =   \sib^{\ald \al}_\mu \Phi_{0\al} ,& \lb{ssp1}\\
\Phi_{1\mu}^{\star\al}   =  \Phi^{\star}_{0 \ald}\sib^{\ald \al}_\mu  ,&
\Phi_{1\ald}^{\star\mu}   =   \Phi^{\star \al}_0 \si_{\al \ald}^\mu  ,&\\
\Psi^{\mu \nu}_{2\al}   =  \Psi_2^\beta\si_{\beta \al}^{\mu \nu}  ,&
\Psi_{2\mu \nu}^{\ald}   =  \Psi_{2\betd}\sib^{\betd \ald}_{\mu \nu}  ,&\\
\phi_{2\mu \nu}^{\al}    =  \si^{\al \beta }_{\mu \nu} \phi_{2\beta},& 
\phi^{\mu \nu}_{2\ald}   =  \sib_{\ald \betd }^{\mu \nu} \phi_2^\betd ,&
 \lb{ssp4}
\eeqa
where $\si_0$ is $(-1)$identity, 
$\si_{1,2,3}$ are the Pauli matrices and
\[
 \si^{\mu \nu \beta}_\al  =
\fr{1}{4} (
\si_{\al \ald}^\mu
\sib^{\nu \ald \beta} -
\si_{\al \ald}^\nu
\sib^{\mu \ald \beta} ) .
\]
Plugging (\ref{ssp1})--(\ref{ssp4})
into (\ref{spsi1}) leads to
\be
\lb{1bl}
S_1=\fr{1}{2}<- i\Psi_0^\al \CD_{\al\ald}\Phi_0^{\ald}
+\Psi_0^{\al} [\etat ,\Psi_{0\al}^{\star}] 
-\fr{3i}{2}\Phi_{0\ald}^{\star} \CD^{\ald\al}\phi_{2\al}
+2\Phi^\star_{0\ald} [ \etat , \Phi_0^\ald ]
-12\Psi_2^\al [ \etat ,\phi_{2\al}] >.
\ee
Similarly, one can show that employing
(\ref{ssp1})--(\ref{ssp4})
in  (\ref{spsi2}) yields
\be
\lb{2bl}
S_2=\fr{1}{2} <
- i\Psi_{0\ald} \CD^{\ald\al}\Phi_{0\al}
+\Psi_{0\ald} [ \etat ,\Psi_{0}^{\star\ald}] 
-\fr{3i}{2}\Phi_0^{\star \al} \CD_{\al\ald}\phi_2^\ald
+2\Phi^{\star\al}_0 [ \etat ,\Phi_{0\al} ]
-12\Psi_{2\ald} [ \etat , \phi_2^\ald ] >.
\ee

By analyzing ghost numbers of component fields 
one can observe that 
the natural identifications are
\beqa
\Psi_{0\al}  \equiv  \omo_\al , &
\Psi^\star_{0\al}  \equiv  \omo^\star_\al ,&  \lb{ide}\\
\Phi_{0\ald}  \equiv  \omob_\ald ,&
\Phi_{0\ald}^{\star}  \equiv  \omob^{\star}_\ald .& \lb{ide1}
\eeqa
Here $\omo^\star$ and $\omob^\star$ are given in accord with
(\ref{omo}) and (\ref{omob}),
by the antifields $A^\star ,\chi^\star$ of 
$A,\chi ,$ as 
\beqa
\omo_\al^\star  =  \teb^\ald A_{\al\ald}^\star +ic\sqrt{3}\teb^2\chi_\al^\star 
+\fr{2ic}{\sqrt{3}}\ep_{\al \bet}\te^\bet \teb^\betd \chib_\betd^\star 
+\fr{i}{2}\teb^2\te_\bet \rD^{\betd \bet}A_{\betd\al}^\star
+\fr{c\sqrt{3}}{5}\te^2\teb^2 \rD_{\al\ald}\chib^{\star \ald} , \lb{omos}\\
\omob_\ald^\star   =  \te^\al A_{\al\ald}^\star -ic\sqrt{3}\te^2c\chib_\ald^\star 
-\fr{2ic}{\sqrt{3}}\ep_{\ald \betd}\te^\bet \teb^\betd \chi_\bet^\star 
 -\fr{i}{2}\te^2\teb_\betd \rD^{\betd \bet}A_{\ald\bet}^\star
+\fr{c\sqrt{3}}{5}\te^2\teb^2\rD_{\al\ald}\chi^{\star\al} . \lb{omobs}
\eeqa
Because of the fact that
antifields 
$A^\star ,\chi^\star$ 
do not have the same dimensions of the original fields
$\chi , A,$ 
we introduced the constant $c$ whose dimension is 
\be
\lb{m}
[c]=[A^\star ]-[\chi^\star ]+1/2=1,
\ee 
so that, the components of $\omo^\star$ possess the same
dimension. Although $c$ will appear in 
the supersymmetry transformations
of the antifields, the final action will be 
independent of it.
 
Putting (\ref{1bl}) and 
(\ref{2bl}) together results in the full action
which after
plugging (\ref{ide})--(\ref{ide1}),
yields
\beqa
S&=&S_1+S_2
= \fr{i}{2}< \int d^4\tau 
\left[ \Psi_{\tau \ald} 
\CD_\tau \Phi_\tau^\ald +
\Psi_\tau^\al \CD_\tau 
\Phi_{\tau \al} \right] > \nonumber \\
&=& S_0+
<-\fr{3i}{4}\omob_{\ald}^{\star} \CD^{\ald\al}\phi_{2\al}
-\fr{3i}{4}\omo^{\star \al} \CD_{\al\ald}\phi_2^\ald
+\fr{1}{2}\omo^{{\star}\al} [\etat ,\omo_{\al}] 
+\fr{1}{2}\omob^\star_{\ald} [ \etat , \omob^\ald ]  \nonumber \\
 & & -6\Psi_2^{ \al} [ \etat ,\phi_{2\al}]
-6\Psi_{2 \ald} [ \etat ,\phi_2^{\ald}] >.  \lb{s12}
\eeqa
To show that this leads to  the minimal
 solution of the BV master equation for supersymmetric
Yang--Mills theory,
we should specify the related component fields.

Ghost superfield $\etat$  should depend
linearly on $\eta .$ Thus, we adopt the definition
\be
\lb{etat}
\etat =\fr{2}{c}\eta 
+\fr{3}{2}\te^2 \eta +\fr{3}{2}\teb^2 \eta ,
\ee
where the constant $c$ has already been introduced in  
(\ref{omos})--(\ref{m}).
Because of not having a spinor ghost field 
a super partner for $\eta$ does not exist.
Hence, $\etat$ is not a representation of  supersymmetry 
algebra.
Obviously, it would be nice to introduce a spinor ghost
field without altering the original supersymmetry
and the number of physical fields to
achieve an extended
solution of BV master equation which possesses
 $N=1$ Yang--Mills supersymmetry.
Although the action (\ref{s12}) may be the appropriate
beginning point 
to find  $N=1$ 
supersymmetric 
solution of BV master equation,
obtaining it is out of the scope of this work. We would like to
obtain
the minimal solution (\ref{solu}) which  possesses BRST symmetry
which is Grassmannian, but it is not invariant under the
$N=1$ supersymmetry of the action (\ref{soo}). Hence, the
ghost superfield (\ref{etat}) reflects the fact that
when obtaining the minimal solution (\ref{solu})
from the action written in terms of the superfields
(\ref{s12}), we should somehow 
break the starting supersymmetry. 

The superfields $\phi_2$ and $ \Psi_2$ possessing
ghost numbers $1$ and $-2$ should be 
proportional to  $\etat$
and its antisuperfield $\etat^\star$ obtained
from the former by replacing the ghost field
$\eta$ with its antifield $\eta^\star$. 
Hence, we define, by suitable normalizations,
\be
\lb{cof}
\phi^\al_2  =\fr{8}{9}\etat \te^\al, \
\phi^\ald_2  = \fr{8}{9}\teb^\ald \etat ; \
\Psi^\al_2  = \fr{3c^2}{8^3} \phi^{\star\al}_2
=\fr{-c^2}{3\cdot 8^2}\te^\al\etat^\star ,
\Psi^\ald_2  = \fr{3c^2}{8^3} \phi^{\star\ald}_2
=\fr{c^2}{3\cdot 8^2}\teb^\ald\etat^\star .
\ee

By performing the integrals over 
the anticommuting variables 
$\te_\al,\ \teb_\ald $ 
we get
\be
\lb{solu}
S=S_0  +
\int d^4x \left\{
-iA^{\star\ald\al}\rD_{\al\ald}\eta
+\chi^{\star \al} [ \eta , \chi_\al ] 
+ \bar{\chi}_\ald^\star [ \eta , \bar{\chi}^\ald ]
+\fr{1}{2}\eta^\star [ \eta , \eta ] \right\}. 
\ee
$S_0$ is given in (\ref{soo}).
Indeed, (\ref{solu}) is the minimal solution of the 
 BV master equation for $N=1$ supersymmetric Yang--Mills
theory.

\section{Discussions}

The main achievements of this work are to write
either $N=1$ supersymmetric action (\ref{soA})
or its BV--quantized action  (\ref{s12}) in very simple forms.
Thus, studying some geometrical
aspects like BRST cohomology should be easier. 
Moreover, simplifications should arise in
calculations of the related path integrals.

This work is 
constituting  an  example to  the
vast applicability
of the generalized fields or $\tau$--fields formalisms:
Although both of the methods can be applied to actions
which are seemingly first order in derivatives, this dependence 
is only in shape. Therefore, applicability of these approaches is not
restricted to the theories whose actions can be written as  first order in
derivatives. 

Theories with higher supersymmetry are difficult to handle. 
Hence, generalizations of our results
to $N>1$ theories would  be  extremely useful for explicit
calculations and to reveal their geometrical aspects.

\newpage

\begin{center}
{\bf Acknowledgment}
\end{center}

I thank Kayhan \"{U}lker for discussions,
checking some calculations and
for his helps regarding the superspace notations.

\end{document}